\documentclass[aps,preprint,nofootinbib,eqsecnum]{revtex4}
\usepackage{amsmath}
\usepackage{amsfonts}
\usepackage{amssymb}
\usepackage{graphicx}
\usepackage{amscd}

 \allowdisplaybreaks

\input{tcilatex}

\begin{document}

\begin{flushleft}
USCHEP/0308ib5
\hfill hep-th/0312147 \\
UCB-PTH-03/33 \hfill CERN-CH/2003-256 \\
LBNL-54091
\end{flushleft}

\title{A Hidden Twelve-Dimensional SuperPoincar\'{e} Symmetry\\
In Eleven Dimensions}
\author{Itzhak Bars}
\affiliation{Department of Physics, USC, Los Angeles, CA 90089-0484, USA \\
Theory Division, CERN, CH-1211 Geneva 23, Switzerland}
\author{Cemsinan Deliduman }
\affiliation{Feza G\"{u}rsey Institute, \c{C}engelk\"{o}y 81220, \.{I}stanbul, Turkey}
\author{Andrea Pasqua and Bruno Zumino}
\affiliation{Lawrence Berkeley National Laboratory, 1 Cyclotron Rd., Berkeley, CA 94720,
USA}

\begin{abstract}
First, we review a result in our previous paper, of how a
ten-dimensional superparticle, taken off-shell, has a hidden
eleven-dimensional superPoincar\'e symmetry. Then, we show that
the physical sector is defined by three first-class constraints
which preserve the full eleven-dimensional symmetry. Applying the
same concepts to the eleven dimensional superparticle, taken
off-shell, we discover a hidden twelve dimensional superPoincar\'e
symmetry that governs the theory.
\tableofcontents
\end{abstract}

\maketitle



\section{From Ten To Eleven Dimensions}

In \cite{BDPZ}, we showed that the quantum algebra of a ten-dimensional
superparticle, taken off shell, contains a nonlinear realization of the
eleven-dimensional superPoincar\'e algebra, with some additional
constraints. In this section, we review the procedure outlined in \cite{BDPZ}
and we write the constraints in a fully covariant fashion.

As a starting point, one can take for instance the Brink-Schwarz action \cite%
{brinkschwarz} for a ten-dimensional massless superparticle. The phase space
of such a particle is spanned by the canonical variables $x^\mu$, $%
\theta^\alpha$, and their respective momenta $p_\mu$, $\pi_\alpha$, where $x$%
, $p$ are vectors in ten dimensions and $\theta$, $\pi$ belong each to a
Majorana-Weyl representation of the ten-dimensional Clifford algebra.

The straightforward quantization of this phase space is impeded by the
presence of constraints, namely
\begin{equation}
p^2=0,\quad d_{\alpha}\equiv \pi _{\alpha}-\left( \mathord{\not\mathrel{p}}%
\theta \right) _{\alpha}=0.
\end{equation}
As in \cite{BDPZ}, we make the choice of ignoring the first constraint,
because we want to describe the quantum mechanics of a particle off-shell.
We retain instead the fermionic constraints. Off-shell, they are
second-class and can be treated by an extension of the Dirac quantization
method \footnote{%
There are some subtleties in extending Dirac's method to superspace, but
they can be overcome for a class of algebras, of which ours is one.}\cite%
{casalbuoni}, \cite{brinkschwarz}: first we reduce by half the number of
fermionic degrees of freedom using the constraints ($\pi _{\alpha }=\left( %
\mathord{\not\mathrel{p}}\theta \right) _{\alpha}$), then we compute the
Dirac brackets for the remaining variables and finally we quantize the Dirac
brackets.

It is interesting to express the remaining fermionic generators in terms of
the supersymmetry generators $Q_{\alpha }\equiv\pi _{\alpha} +\left( %
\mathord{\not\mathrel{p}}\theta \right) _{\alpha}= 2\pi_{\alpha}$. If we do
that, the quantum algebra that we obtain is the following
\begin{eqnarray}
\left\{ Q_{\alpha },Q_{\beta }\right\} =2\left( \mathord{\not\mathrel{p}}%
\right) _{\alpha \beta },\quad \left[ Q_{\alpha },p_{\mu }\right] =0,\quad %
\left[ p_{\mu },p_{\nu }\right] =0,  \label{susy} \\
\left[ x^\mu,p_\nu \right] =i\delta^\mu_\nu , \quad \left[ Q_{\alpha
},x^{\mu }\right] =-\frac{i}{2}\left( \gamma ^{\mu }\mathord{\not\mathrel{p}}%
^{-1}Q\right) _{\alpha },\quad  \label{noncomm1} \\
\left[ x^{\mu },x^{\nu }\right] =-\frac{1}{16p^{4}}Q\left\{ \gamma ^{\mu \nu
},\mathord{\not\mathrel{p}}\right\} Q.\qquad \qquad \qquad  \label{noncomm2}
\end{eqnarray}
This quantum algebra in ten dimensions is free of constraints. It contains
the ten-dimensional supertranslations and has an interesting
noncommutativity in the spacetime coordinates. Its consistency can be
verified \cite{BDPZ} by checking that all Jacobi identities are verified.

Next, we consider the following elements of the algebra:
\begin{eqnarray}
J^{\mu }\equiv \left( -p^{2}\right) ^{\frac{1}{4}}x^{\mu }\left(
-p^{2}\right) ^{\frac{1}{4}},  \label{fromJtoX} \\
J^{\mu \nu }\equiv \left( x^{\mu }p^{\nu }-x^{\nu }p^{\mu }\right) +S^{\mu
\nu },  \label{J} \\
\tilde{Q}_{\dot{\alpha}}\equiv \left( -p^{2}\right) ^{-\frac{1}{2}}\left( %
\mathord{\not\mathrel{p}}Q\right) _{\dot{\alpha}},
\end{eqnarray}
where
\begin{equation}
S^{\mu \nu} \equiv \frac{-i}{16p^{2}}Q\left\{ \gamma ^{\mu \nu },%
\mathord{\not\mathrel{p}}\right\} Q.  \label{spin}
\end{equation}

>From them, we can construct the generators of an eleven-dimensional
superPoincar\'{e} algebra \footnote{%
Whenever square roots appear it is understood that both signs may occur in
front of them, so that in particular $P^{10}=\pm \sqrt{-p^{2}}$ spans the
whole range of momentum in the extra dimension. To avoid cluttering our
notation we omit the extra $\pm$.} as follows \cite{BDPZ}
\begin{eqnarray}
P^{M} &=&(p^{\mu },P^{10}\equiv \sqrt{-p^{2}}),  \label{bigP} \\
J^{MN} &=&\left( J^{\mu \nu },J^{\mu ,10}\equiv J^{\mu }\right) ,
\label{bigJ} \\
Q_{A} &=&\left( Q_{\alpha },\tilde{Q}_{\dot{\alpha}}\equiv \frac{(%
\mathord{\not\mathrel{p}}Q)_{\dot{\alpha}}}{\sqrt{-p^{2}}}\right) ,
\label{bigQ}
\end{eqnarray}%
where the indices $M$ and $N$ now range from 0 to 10, and the index $A$ is
the index of a Majorana representation of the Clifford algebra in eleven
dimensions \footnote{%
We have indicated both the ten-dimensional operators $Q_{\alpha }$ and the
eleven-dimensional operator $Q_{A}$ with the same letter. However, it should
be clear from the context which is which. In particular in the following two
sections all the instances of the letter $Q$ refer to the eleven-dimensional
Majorana spinor.}. It is a matter of straightforward computation to show
that $P$, $J$ and $Q$ indeed satisfy the superPoincar\'{e} algebra in one
more dimension than we started with. We write its commutation and
anticommutation relations at the end of the next section.

Note that this eleven-dimensional algebra is realized nonlinearly in the
original ten-dimensional algebra, but it also contains it, because among its
generators are in particular $J^\mu\equiv J^{\mu,10}$, $p_\mu$ and $Q_\alpha$
which in turn generate the original algebra \footnote{$x^\mu$ can be
expressed in terms of $J^\mu$ by inverting (\ref{fromJtoX}).}. Hence, the
two algebras are actually the same. This apparent paradox is resolved once
we realize that the new algebra is not free. In the next section we describe
the constraints it is subject to.

Note also that so far we have assumed a timelike momentum for the off-shell
particle, so that $\sqrt{-p^2}$ is real. More generally we should allow also
an off-shell spacelike momentum. In that case the extra dimension is
timelike because the momentum in the additional dimension is purely
imaginary and given by $i\sqrt{p^2}$. In the latter case, all our
constructions can be extended and the corresponding formulas can be obtained
by analytic continuation $\sqrt{-p^2}\rightarrow i\sqrt{p^2}$ from the ones
given below. In the following, we will let it be understood that when $\sqrt{%
-p^2}$ is real the extra dimension is spacelike, and when it is imaginary
the extra dimension is timelike.

\section{Constraints}

To begin with, the eleven-dimensional algebra satisfies the constraints
\begin{eqnarray}
P^{M}P^{N}\eta _{MN} &=&0,  \label{massless1} \\
P^{M}\left( \Gamma _{M}\right) _{A}^{~B}Q_{B} &=&0,  \label{massless2}
\end{eqnarray}%
where $\eta _{MN}$ is the Minkowski metric in eleven dimensions, the last
dimension taken to be spacelike, and the $\Gamma ^{M}$ form a representation
of the Clifford algebra in eleven dimensions, whose expression in terms of $%
\gamma _{\alpha \beta }^{\mu }$ matrices (and their antichiral counterpart $%
\gamma _{\dot{\alpha}\dot{\beta}}^{\mu }$) is
\begin{equation}
\Gamma ^{\mu }=\left(
\begin{array}{cc}
0 & \gamma _{\alpha \beta }^{\mu } \\
\gamma _{\dot{\alpha}\dot{\beta}}^{\mu } & 0%
\end{array}%
\right) ,\;\Gamma ^{10}=\left(
\begin{array}{cc}
1 & 0 \\
0 & -1%
\end{array}%
\right) .  \label{gamma11}
\end{equation}%
Specifically, (\ref{massless1}) encodes in the new algebra the definition (%
\ref{bigP}) of $P$ in terms of quantities in the old algebra, and similarly (%
\ref{massless2}) encodes (\ref{bigQ}).

The algebra has also a constraint that encodes the definition (\ref{bigJ}).
It is easy to write it in a ten-dimensional covariant way by combining (\ref%
{fromJtoX}), (\ref{J}) and (\ref{spin}).
\begin{equation}
J^{\mu \nu }-\left( -p^{2}\right) ^{-\frac{1}{4}}\left( J^{\mu }p^{\nu
}-J^{\nu }p^{\mu }\right) \left( -p^{2}\right) ^{-\frac{1}{4}}=-\frac{i}{%
16p^{2}}Q\left\{ \gamma ^{\mu \nu },\mathord{\not\mathrel{p}}\right\} Q,
\label{thirdconstr}
\end{equation}%
This constraint contains a nontrivial relation between the bosonic and the
fermionic parts of the eleven-dimensional algebra. It also allows us to
express $J^{\mu \nu }$ in terms of $J^{\mu }$, $p^{\mu }$ and $Q_{\alpha }$,
making explicit the fact that the constrained eleven-dimensional algebra has
the same number of independent generators as the algebra in ten dimensions.

It is desirable to express (\ref{thirdconstr}) in an eleven-dimensionally
covariant form. For that purpose, first we rewrite every quantity in the
constraint explicitly as a generator of the eleven-dimensional algebra, then
we perform a few algebraic steps and we obtain
\begin{equation}
J^{\mu \nu }P^{10}+J^{\nu 10}P^{\mu }+J^{10\mu }P^{\nu }=\frac{i}{16}\left(
Q\gamma ^{\mu \nu }\tilde{Q}+\tilde{Q}\gamma ^{\mu \nu }Q\right) .
\label{thirdconstr2}
\end{equation}%
The left-hand side of (\ref{thirdconstr2}) can be written as $3\times
W^{10\mu \nu }$, with
\begin{equation}
W^{LMN}\equiv J^{<LM}P^{N>}\equiv \frac{1}{3!}\left( J^{LM}P^{N}\pm \text{%
permutations}\right) \footnote{%
In four dimensions, (\ref{Pauli-Lubanski}) is the dual of the
Pauli-Luba\'nski vector, so that $W^{LMN}$ should be thought of as its
generalization to higher dimensions.}.  \label{Pauli-Lubanski}
\end{equation}%
Here and in the following, the angular brackets indicate complete
antisymmetrization. To rewrite covariantly the right-hand side, we need to
find spinor bilinears with tensorial transformation properties. Let us
define
\begin{equation*}
\bar{Q}\equiv Q^{T}\mathcal{C},\quad \mathcal{C}=\left(
\begin{array}{cc}
0 & 1 \\
-1 & 0%
\end{array}%
\right) .
\end{equation*}%
The matrix $\mathcal{C}$ is chosen so that it satisfies $\mathcal{C}\Gamma
^{M}\mathcal{C}^{\dagger }=-(\Gamma ^{M})^{T}$. Then $\bar{Q}\Gamma
^{M_{1}\dots M_{p}}Q$ \footnote{$\Gamma ^{M_{1}\dots M_{p}}$ indicates the
antisymmetric combination of $p$ gamma matrices, $\Gamma ^{<M_{1}}\dots
\Gamma ^{M_{p}>}$} transforms as an antisymmetric $p$-tensor under the
Lorentz group in eleven dimensions. An explicit computation shows that the
right-hand side of (\ref{thirdconstr2}) is $3\times S^{10\mu \nu }$, with
\begin{equation}
S^{LMN}\equiv -\frac{i}{3\times 16}\bar{Q}\Gamma ^{LMN}Q,  \label{S}
\end{equation}%
so that the third constraint (\ref{thirdconstr}) reads simply $W^{10\mu \nu
}=S^{10\mu \nu }$. Furthermore, the equality holds also for the other
components of $W$ and $S$. This can be checked by explicit computation in
ten-dimensional language and, again, it is a consequence solely of (\ref%
{thirdconstr}). In conclusion, the third constraint (\ref{thirdconstr}) can
be written as
\begin{equation}
\Delta ^{LMN}\equiv W^{LMN}-S^{LMN}=0.  \label{third}
\end{equation}

It should be clear from the previous line of reasoning that not all
components of (\ref{third}) are independent of one another. Indeed the
number of independent components has to be that of the equation (\ref%
{thirdconstr}) from which we started. That number is $\left(%
\begin{array}{c}
10 \\
2%
\end{array}%
\right)$. There is a more elegant and fully covariant way to see that the
number of independent components in (\ref{third}) is indeed $\left(%
\begin{array}{c}
10 \\
2%
\end{array}%
\right)$, and we show it in the next subsection.

Thus, the conclusion of our analysis is that the Hilbert space of a
ten-dimensional superparticle taken off-shell is also the Hilbert space of
the eleven-dimensional superPoincar\'e algebra
\begin{eqnarray}
\left[P^M,P^N \right]=0, \; \left[P^M,Q_A \right]=0, \; \left[P^M,J^{NQ} %
\right]=i \eta^{MQ} P^N -i \eta^{MN}P^Q \\
\left\{Q_A,\bar{Q}_B \right\}=2 \mathord{\not\mathrel{P}}_{AB}, \qquad \left[
J^{MN}, Q_A \right]=\frac{i}{2}(\Gamma^{MN}Q)_A \qquad \\
\left[ J^{MN}, J^{PQ} \right]=i \eta^{MP} J^{NQ} - i \eta^{MQ} J^{NP} +i
\eta^{NQ} J^{MP}-i \eta^{NP} J^{MQ}
\end{eqnarray}
constrained as follows
\begin{equation}
P^2=0, \quad \mathord{\not\mathrel{P}}Q=0, \quad \Delta^{LMN}=0.
\end{equation}%
\label{final}

The first two constraints are well known in the context of the
massless superparticle in eleven dimensions. The last constraint
$\Delta ^{LMN}=0,$ and some additional ones to be discussed in the
next section, are newly realized. As is well known, the spectrum
of quantum states that satisfy the first two constraints is
precisely the supergravity multiplet in eleven dimensions,
consisting of the
metric $g_{MN}$, 3-index antisymmetric tensor $A_{LMN}$, and the gravitino $%
\psi _{A}^{M}$. The last constraint, and the additional ones discussed in
the next section, are also satisfied covariantly by this supermultiplet.

In fact, other than the 11D supergravity multiplet, there are no other
supermultiplets that satisfy these constraints. This can be seen by solving
the constraints explicitly in terms of the 10D unconstrained degrees of
freedom, which correspond to the off-shell 10D superparticle whose quantum
states correspond to the 11D supergravity multiplet but in a 10D notation.
This point can also be understood by solving the constraints in the
lightcone gauge of the eleven dimensional superparticle, which also
indicates the same set of quantum states in a fixed gauge; namely 128 bosons
and 128 fermions consisting of only the SO(9) covariant transverse degrees
of freedom of the supergravity multiplet, $g_{ij}, A_{ijk}, \psi_{a}^i$.
Incidentally, in the context of the 11D superparticle, we should emphasize
that our approach provides a ghost-free SO(9,1) covariant quantization of
the 11D superparticle. This displays more symmetry as compared to the
ghost-free light-cone quantization.

\subsection{Counting Constraints Covariantly}

To count correctly the number of independent constraints in equation (\ref%
{third}), we need to take into account that not all components of that
equation are independent. One way to show that this is the case is to show
that there is a constraint on the constraint. Indeed it can be checked that
\begin{equation}
\Delta^{LMNO}\equiv \Delta^{<LMN}P^{O>}=0,  \label{thirdprime}
\end{equation}
by virtue of the definition of $\Delta^{LMN}$ alone, without using the
condition $\Delta^{LMN}=0$. It holds trivially for the $W$ part of $\Delta$,
while a brief computation is required to show that it holds also for the $S$
part \footnote{%
For instance, one can start with the equations $\bar{Q}\left\{ %
\mathord{\not\mathrel{P}},\Gamma ^{M_{1}\cdots M_{p}}\right\} Q=\bar{Q}\left[
\mathord{\not\mathrel{P}},\Gamma ^{M_{1}\cdots M_{p}}\right] Q=0$, which
follow from the constraint $\mathord{\not\mathrel{P}}Q=0,$ and evaluate the
anticommutator or commutator using the Clifford algebra. From this one can
show that $\bar{Q}\Gamma ^{<M_{1}\dots M_{p-1}}QP^{M_{p}>}=0$ and $\bar{Q}%
\Gamma ^{M_{1}\dots M_{p+1}}QP_{M_{p+1}}=0$ for an arbitrary number of
indices. Applying this to $p=1$ we derive $\bar{Q}Q=0.$ In addition, it is
possible to show that $\bar{Q}\Gamma ^{M_{1}\dots M_{p}}Q$ vanishes by
itself for $p=2,5,6,9,$ which corresponds to the vanishing of the D-brane
charges in the 11D superalgebra. The $p=1,10$ cases are simple $\bar{Q}%
\Gamma ^{M}Q=32P^{M}$, $\bar{Q}\Gamma ^{M_{1}\cdots M_{10}}Q=32\epsilon
^{M_{1}\cdots M_{11}}P_{M_{11}}$, while the remaining cases $p=3,4,7,8$
satisfy the above constraints nontrivially.}. Therefore (\ref{thirdprime})
is an honest constraint on the constraint. Again, not all components of (\ref%
{thirdprime}) are independent of one another and indeed they are subject to
a constraint themselves, namely
\begin{equation}
\Delta^{LMNOP}\equiv \Delta^{<LMNO}P^{P>}=0,  \label{thirddoubleprime}
\end{equation}
and so on.

There is an end to this chain of constraints, because each of the
constraints is completely antisymmetric in its indices and so it can have at
most 11 indices. Incidentally, the operation of adding a power of $P$ and
antisymmetrizing can be thought of as a cohomological operation, akin to
taking the exterior derivative in De Rham cohomology. To count the correct
number of degrees of freedom then, we ought to start from the end. The last
constraint is $\Delta^{L_1\dots L_{11}}=0$. Because of antisymmetry, this
has only $\left(%
\begin{array}{c}
11 \\
11%
\end{array}%
\right)=1$ independent components and it constraints the previous equation
in the chain $\Delta^{L_1\dots L_{10}}=0$, so that the latter has only $%
\left(%
\begin{array}{c}
11 \\
10%
\end{array}%
\right) -\left(%
\begin{array}{c}
11 \\
11%
\end{array}%
\right)= 10-1$ independent components. These are the number of components
that should be subtracted from the number of components in the previous
equation yet, and so on backwards along the chain. Consequently, the number
of independent components in the third constraint (\ref{third}) must be
\begin{equation}
\left(%
\begin{array}{c}
11 \\
3%
\end{array}%
\right)- \left(%
\begin{array}{c}
11 \\
4%
\end{array}%
\right)+ \left(%
\begin{array}{c}
11 \\
5%
\end{array}%
\right)- \dots - \left(%
\begin{array}{c}
11 \\
10%
\end{array}%
\right)+ \left(%
\begin{array}{c}
11 \\
11%
\end{array}%
\right)= \left(%
\begin{array}{c}
10 \\
2%
\end{array}%
\right),  \notag
\end{equation}
as expected.

\section{The Role Of Supersymmetry}

Of the three constraints that the eleven-dimensional algebra is subject to,
the third appears somewhat peculiar, especially on account of the
coefficient entering the definition of $S^{LMN}$. We found that some light
can be shed by examining the transformation properties of the constraints
under supersymmetry.

Before we do that, let us begin with a premise. In the previous section, we
expressed the constraints on the eleven-dimensional algebra as the vanishing
of tensorial and spinorial quantities in eleven-dimensions. As such, the
constraints are automatically consistent with the Lorentz part of the
algebra, in the sense that their variation under Lorentz transformations
\footnote{%
We define the Lorentz transformations as $\delta_J (\cdot)\equiv \left[
J^{RS}, \cdot \right]$. Similarly $\delta_P (\cdot)\equiv \left[ P^R, \cdot %
\right]$ and $\delta_Q (\cdot)\equiv \left[ Q_A, \cdot \right]$ for bosons, $%
\equiv \left\{ Q_A, \cdot \right\}$ for fermions.} vanishes once we impose
the constraints themselves. More specifically,
\begin{equation}
\delta_J P^2 = 0,\quad \delta_J (\mathord{\not\mathrel{P}}Q) = \frac{i}{2}
\Gamma^{MN} \mathord{\not\mathrel{P}}Q,\quad \delta_J \Delta^{LMN}= i \eta
^{RL}\Delta^{SMN}+ \dots,  \label{deltaJ}
\end{equation}
and these variations are zero modulo $\mathord{\not\mathrel{P}}Q,
\Delta^{LMN}$.

A different way to put it, is that if we represent the algebra on a Hilbert
space of states, the states that satisfy the constraints are invariant under
the Lorentz subalgebra. The same holds for the translations, because all
constraints commute with $P$.
\begin{equation}
\delta_P P^2 = 0, \quad \delta_P (\mathord{\not\mathrel{P}} Q) = 0, \quad
\delta_P \Delta^{LMN} = 0.  \label{deltaP}
\end{equation}

The next natural step is to check what happens with the supersymmetry
transformations. We find the following
\begin{equation}
\delta_Q (P^2)=0,\quad \delta_Q (\mathord{\not\mathrel{P}}Q)_B = 2 P^2
\mathcal{C}_{AB}, \quad \delta_Q ( \Delta ^{LMN} )= -\frac{i}{12}
(\Gamma^{LMN}\mathord{\not\mathrel{P}}Q)_A.  \label{susyconstr}
\end{equation}
So again we find that the supersymmetry transformations vanish once we
assume the constraints to hold, and therefore that the states that satisfy
the constraints are invariant under the supersymmetry transformations as
well. Hence, the constraints are such that they preserve the full symmetry
of the superPoincar\'e algebra. In addition, the commutators (or
anticommutators) of the constraints with one another can be computed using (%
\ref{deltaJ}), (\ref{deltaP}) and (\ref{susyconstr}). One finds that they
vanish modulo the constraints themselves (but with coefficients that depend
on the dynamical variables). In other words, the constraints are first
class. They generate some supergroup of transformations.

We should point out that the value of $\delta_Q ( \Delta ^{LMN} )$ depends
critically on the choice of coefficient for $S^{LMN}$ in (\ref{S}). With a
different coefficient, there would be residual pieces which are not
proportional to any of the constraints. However, when the coefficient is
chosen to be precisely as in (\ref{S}), a cancellation occurs between $%
\delta_Q (W^{LMN})$ and some terms in $\delta_Q S^{LMN}$ and the only term
left is the one given in (\ref{susyconstr}).

More importantly, we should note that the interplay of the constraints is
more interesting for supersymmetry transformations, because now the
variation of the third constraint $\delta _{Q}(\Delta ^{LMN})$ vanishes only
modulo the second constraint $\mathord{\not\mathrel{P}}Q$, and similarly $%
\delta _{Q}(\mathord{\not\mathrel{P}}Q)$ vanishes only modulo $P^{2}$. In
other words, $\Delta ^{LMN}=0$ is consistent with supersymmetry only if we
also require $\mathord{\not\mathrel{P}}Q=0$. Similarly for $%
\mathord{\not\mathrel{P}}Q$ and $P^{2}$.

Let us also mention that in four dimensions a generalization of
the Pauli-Luba\'{n}ski vector\ was
discussed in \cite{S}\cite{PT}. It was given as $C_{a}=W_{a}-\frac{i}{8}%
\bar{Q}\gamma _{a}\gamma _{5}Q$, where
$W_{a}=\frac{1}{2}\varepsilon _{abcd}p^{a}J^{cd},$ and the Latin
indices are four-dimensional space-time indices. If we specialize
our off-shell superparticle approach to four dimensions (with
hidden five dimensional symmetry), our five dimensional
$\Delta^{LMN}$ has a four dimensional component $\Delta
^{lmn}\sim\varepsilon ^{lmna}\Delta _{a}$ which we can attempt to
compare to $C_{a}$. We find that $\Delta _{a}$ is different than
$C_{a}$ by including an additional crucial term,
$\Delta_{a}=C_{a}+\frac{i}{8p^{2}}p_{a}\bar{Q}
\mathord{\not\mathrel{p}}\gamma _{5}Q$. The supersymmetry
variation of $C_{a} $ is $\delta _{Q}C_{a}=-\frac{i}{2}\gamma
_{5}Qp_{a}$. However, the supersymmetry variation of $\Delta _{a}$
is $\delta _{Q}\Delta _{a}=0.$ For this reason the constraint
$\Delta _{a}=0,$ or more generally the five dimensional $\Delta
^{LMN}=0$ can be imposed without breaking supersymmetry in five
dimensions. In more general representations where $\Delta _{a}\neq
0, $ we note that $\Delta _{a}$ commutes with $p_{b}$ and $\Delta
^{2}$ commutes also with the Lorentz generators. So $\Delta ^{2}$
is a Casimir invariant of the full superPoincar\'e algebra in four
dimensions. For comparison to \cite{S}\cite{PT} one may also
construct from the $\Delta _{a}$ the tensor $C_{ab}\equiv
p_{a}\Delta _{b}-p_{b}\Delta _{a}$ which coincides with
$p_{a}C_{b}-p_{b}C_{a}.$ Then $C_{ab}C^{ab}$ is also a Casimir for
the full algebra related to $\Delta^2$. The eigenvalue of
$C_{ab}C^{ab}$ is proportional to $Y(Y+1)$ where $Y$ is integer or
half-integer. $Y$ was called called ``superspin'' in
\cite{S}\cite{PT}.

\section{From Eleven to Twelve Dimensions}

We repeat the reasoning of the previous sections by taking as the starting
point the off-shell eleven dimensional superparticle. The dynamical quantum
operators of interest are the 11-component vectors ($X^{M},P_{M}$) and the
supercharge $Q_{A}$ which is a 32-component spinor in eleven dimensional
spacetime. By following the same procedure, the nonlinear quantum algebra
that we obtain has a similar form to the ten-dimensional one
\begin{eqnarray}
\left\{ Q_{A},Q_{B}\right\} &=&-2\left( \mathord{\not\mathrel{P}}\mathcal{C}%
\right) _{AB},\quad \left[ Q_{A},P_{M}\right] =0,\quad \left[ P_{M},P_{N}%
\right] =0,  \label{11D1} \\
\left[ X^{M},P_{N}\right] &=&i\delta _{N }^{M },\quad \left[ Q_{A},X^{M}%
\right] =-\frac{i}{2}\left( \Gamma ^{M}\mathord{\not\mathrel{P}}%
^{-1}Q\right) _{A},\quad  \label{11D2} \\
\left[ X^{M},X^{N}\right] &=&-\frac{1}{16P^{4}}\bar{Q}\left\{ \Gamma ^{M N },%
\mathord{\not\mathrel{P}}\right\} Q.\qquad \qquad \qquad  \label{11D3}
\end{eqnarray}%
Note that we are now using the 32$\times $32 gamma matrices $\Gamma ^{M}$
given above$.$ We inserted explicitly the charge conjugation matrix $%
\mathcal{C}$ which satisfies $\mathcal{C}^{-1}{\Gamma }^{M}{C=-}\left({%
\Gamma }^{M}\right) ^{T},$ and have defined $\bar{Q} \equiv Q^{T}\mathcal{C}.
$ The matrices $\left( \Gamma ^{M}\mathcal{C}\right) _{\alpha \beta }$, $%
\left( \Gamma ^{MN}\mathcal{C}\right) _{\alpha \beta }$, $\left( \Gamma
^{M_{1}\cdots M_{5}}\mathcal{C}\right) _{\alpha \beta }$ are 32$\times $32
symmetric, and $\left({\Gamma }^{MNL}\mathcal{C}\right) _{\alpha \beta }$, $%
\left({\Gamma }^{MNLK}\mathcal{C}\right) _{\alpha \beta }$ are 32$\times $32
antisymmetric. This algebra has no constraints. The first line is the
standard eleven-dimensional superPoincar\'{e} algebra, and the rest is a new
nonlinear extension for the case of the off-shell superparticle\footnote{%
The $x^{M}$ become commutative on shell, since then $\mathord{\not%
\mathrel{p}}Q=0$ for the massless superparticle.}. The consistency of this
algebra can be verified as in \cite{BDPZ} by checking that all Jacobi
identities hold.

Next, as before, we consider the following elements of the algebra:
\begin{eqnarray}
J^{M} &\equiv &\left( -P^{2}\right) ^{\frac{1}{4}}X^{M}\left( -P^{2}\right)
^{\frac{1}{4}}, \\
J^{MN} &\equiv &\left( X^{M}P^{N}-X^{N}P^{M}\right) +S^{MN}, \\
\tilde{Q}_{A} &\equiv &\left( -P^{2}\right) ^{-\frac{1}{2}}\left( %
\mathord{\not\mathrel{P}}Q\right) _{A},
\end{eqnarray}%
where
\begin{equation}
S^{MN}\equiv \frac{-i}{16P^{2}}\bar{Q}\left\{ \Gamma ^{MN},%
\mathord{\not\mathrel{P}}\right\} Q.
\end{equation}%
Note that $\tilde{Q}_{A}$ and $Q_{A}$ are both in the 32-component spinor
representation, unlike the ten-dimensional case where $\left( Q_{\alpha },
\tilde{Q}_{\dot{\alpha}}\right) $ were in different representations, namely $%
\left( 16,16^{\ast}\right)$. Therefore, we will use an additional index $%
i=1,2$ to identify $Q_{A}^{i}=\left( Q_{A},\tilde{Q}_{A}\right) $ as two
supercharges that belong to a $N=2$ supersymmetry in 11-dimensions. These
two supercharges satisfy the SO$\left( 10,1\right) \times $ SO$\left(
2\right)$ covariant constraint
\begin{equation}
\left( \mathord{\not\mathrel{P}}Q^{i}\right) _{A}-\left( -P^{2}\right) ^{%
\frac{1}{2}}\varepsilon ^{ij}Q_{A}^{j}=0,  \label{c1}
\end{equation}%
and $S^{MN}$ takes the SO$\left( 2\right) $ invariant form%
\begin{equation}
S^{MN}= - \frac{i}{16\sqrt{-P^{2}}}\bar{Q}^{i}\Gamma ^{MN}{Q}^{j}\varepsilon
_{ij},~
\end{equation}%
where $\varepsilon^{ij}$ is antisymmetric and $\varepsilon^{12}=
-\varepsilon_{12}=+1$. The nonlinear algebra above may now be rewritten as a
nonlinear extension of the $N=2$ eleven dimensional superPoincar\'{e}
algebra consistent with SO$\left( 2\right)$
\begin{eqnarray}
\left\{ Q_{A}^{i},Q_{B}^{j}\right\} &=&-2\delta ^{ij}\left( %
\mathord{\not\mathrel{P}}\mathcal{C}\right) _{AB}-2\varepsilon ^{ij}\left(
-P^{2}\right) ^{\frac{1}{2}}\mathcal{C}_{AB},\quad \left[ Q_{A}^{i},P_{M} %
\right] =0,\quad \left[ P_{M},P_{N}\right] =0, \\
\left[ J^{M},P_{N}\right] &=&i\left( -P^{2}\right) ^{\frac{1}{2}}\delta
_{N}^{M},\quad \left[ J^{M},Q_{A}^{i}\right] =-\frac{i}{2}\varepsilon
^{ij}\left( \Gamma ^{M}Q^{j}\right) _{A},\quad \left[ J^{M},J^{N}\right]
=iJ^{MN}.\qquad \qquad
\end{eqnarray}%
The $J^{MN}$ which was given above in terms of $X^{M},$ is rewritten in
terms of $J^{M},P^{M},$ and $Q_{A}^{i}$ in the SO$\left( 10,1\right) \times $
SO$\left( 2\right) $ covariant notation as
\begin{equation}
J^{MN}=\left( -P^{2}\right) ^{-\frac{1}{4}}\left(
J^{M}P^{N}-J^{N}P^{M}\right) \left( -P^{2}\right) ^{-\frac{1}{4}}+\frac{i}{16%
\sqrt{-P^{2}}}\bar{Q}^{i}\Gamma^{MN}Q^{j}\varepsilon _{ij}~.
\label{JMNconstraint}
\end{equation}%
By using the nonlinear algebra above, it is straightforward to show that $%
J^{MN}$ satisfies the standard Lorentz algebra in 11-dimensions and is the
generator of 11D Lorentz transformations for all the 11D vectors and spinors
that have appeared so far above.

It is also possible to construct the generators of a twelve-dimensional
superalgebra from the unconstrained operators $P^{M},J^{M},Q_{A}^i$, as
follows. We construct the twelve dimensional operators as
\begin{eqnarray}
P^{m} &=&(P^{M},P^{11}\equiv \sqrt{-P^{2}}),  \label{bigP12} \\
J^{mn} &=&\left( J^{MN},J^{M,11}\equiv J^{M}\right) ,  \label{bigJ12} \\
q_{a} &=&\frac{1}{\sqrt{2}}\left( Q_{A}^{1}+iQ_{A}^{2}\right) =\frac{1}{%
\sqrt{2}}\left( \left( 1+i\left( -P^{2}\right) ^{-\frac{1}{2}}%
\mathord{\not\mathrel{P}}\right) Q\right) _{A},  \label{q12} \\
\bar{q}_{\dot{a}} &=&\frac{1}{\sqrt{2}}\left( \bar{Q}_{A}^{1}-i\bar{Q}%
_{A}^{2}\right) =\frac{1}{\sqrt{2}}\left( \bar{Q}\left( 1+i\left(
-P^{2}\right) ^{-\frac{1}{2}}\mathord{\not\mathrel{P}}\right) \right) _{A}.
\label{qbar12}
\end{eqnarray}%
The indices $m$ and $n$ now range from 0 to 11, and the indices $a,\dot{a}$
denote the complex spinors of SO$\left( 11,1\right) $ which are \textbf{32}
and \textbf{32}$^{\ast }.$ \footnote{%
Note that if the off-shell momentum in eleven dimensions is spacelike, then
we would obtain two \textit{real} chiral spinors in twelve dimensions
belonging respectively to \textbf{32} and \textbf{32'} representations of SO$%
\left( 10,2\right)$. Indeed the analytic continuations of $q_a$ and $\bar{q}%
_{\dot{a}}$ for $P^2$ positive are both real.} It can then be shown that the
$N=2$ nonlinear algebra in 11 dimensions now takes the form of the linear
12-dimensional superPoincar\'{e} algebra given below%
\begin{eqnarray}
\left\{ q_{a},\bar{q}_{\dot{b}}\right\} &=&2\left( \mathord{\not\mathrel{P}}%
\right) _{a\dot{b}},\quad \left[ q_{a},P_{m}\right] =0,\quad \left[
P_{m},P_{n}\right] =0, \\
\left[ J^{mn},P^{l}\right] &=&i\left( \Sigma ^{mn}\right)
_{k}^{l}P^{k},\quad \left[ J^{mn},q_{a}\right] =\frac{i}{2}\left( \Gamma
^{mn}q\right) _{a},\quad \left[ J^{mn},J^{kl}\right] =if_{rs}^{mn,kl}J^{rs}.%
\qquad \qquad
\end{eqnarray}%
where the first line is the standard SUSY algebra in twelve dimensions with $%
\left( \Gamma ^{m}\right) _{a\dot{b}}=\left( \left( \Gamma ^{M}\right)
_{AB},-i\delta _{AB}\right) $, while the second line contains the expected
commutation properties of the SO$\left( 11,1\right) $ generator $J^{mn},$
with
\begin{eqnarray}
\left( \Sigma ^{mn}\right) _{k}^{l} &=&\eta ^{ml}\delta _{k}^{n}-\eta
^{nl}\delta _{k}^{m}, \\
f_{rs}^{mn,kl} &=&\left[ \eta ^{mk}\delta _{r}^{n}\delta _{s}^{l}-\left(
k\leftrightarrow l\right) \right] -\left[ m\leftrightarrow n\right], \\
\left( \Gamma ^{mn}\right) _{ab} &=&\left( \left( \Gamma ^{MN}\right)
_{AB},~\left( \Gamma ^{M,11}=i\Gamma ^{M}\right) _{AB}\right).
\end{eqnarray}%
The antichiral counterparts of the matrices $\left( \Gamma ^{m}\right) _{a%
\dot{b}}$ and $\left( \Gamma ^{mn}\right) _{ab}$ above are respectively
\begin{eqnarray}
\left( \Gamma ^{m}\right) _{\dot{a}b}&=&\left( \left( \Gamma ^{M}\right)
_{AB},i\delta _{AB}\right), \\
\left( \Gamma ^{mn}\right) _{\dot{a}\dot{b}} &=&\left( \left( \Gamma ^{MN}
\right)_{AB},~\left( \Gamma ^{M,11}=-i\Gamma ^{M}\right) _{AB}\right).
\end{eqnarray}
This algebra is subject to the three 12D covariant constraints%
\begin{equation}
P^{m}P_{m}=0,\;P_m (\Gamma^m)_{\dot{a}b} q_b=0,\;\Delta ^{mnl}=0,
\label{12Dconstraints}
\end{equation}%
where the last two are a 12D covariant rewriting of the 11D constraints in
Eqs.(\ref{c1},\ref{JMNconstraint}). The tensor $\Delta ^{mnl}$ is defined by
$\Delta ^{lmn}=W^{lmn}-S^{lmn},$ where $W^{lmn}$ and $S^{lmn}$ are the
generalized Pauli-Luba\'{n}ski and spin tensors in 12D
\begin{equation}
W^{lmn}\equiv J^{<lm}P^{n>}\equiv \frac{1}{3!}\left( J^{lm}P^{n}\pm \text{%
permutations}\right) ,\;S^{lmn}\equiv -\frac{i}{3\times 16}\bar{q}\Gamma
^{lmn}q
\end{equation}%
We see that the constraint in Eq.(\ref{JMNconstraint}) corresponds to
\begin{equation}
J^{\mu \nu }P^{11}+J^{\nu 11}P^{\mu }+J^{11\mu }P^{\nu }-\frac{i}{16}%
\varepsilon ^{ij}\bar{Q}^{i}\gamma ^{\mu \nu }Q^{j}=3\Delta ^{11mn}=0.
\end{equation}
The number of independent components in the tensor $\Delta^{lmn}$ can be
computed covariantly in 12 dimensions as in the previous section,
\begin{equation}
\sum_{k=3}^{12}\binom{12}{k}\left( -1\right) ^{k+1}=55=\left( 1-1\right)
^{12}-\left[ -1+\binom{12}{1}-\binom{12}{2}\right]
\end{equation}
This is the same number of components as in Eq.(\ref{JMNconstraint}), namely
$11\times 10/2=55$.

The supermultiplet of the quantum states that provide a representation of
the constrained 12D superPoincar\'{e} algebra can be easily computed in the
SO(10) covariant lightcone gauge. After solving all the constraints
explicitly, the degrees of freedom reduce to $x^i,p^i,\chi^a$ where the
SO(10) vectors $x^i,p^i$ are canonical and the 32-components of $\chi^a$
(two SO(10) spinors) satisfy the Clifford algebra $\{\chi^a,\chi^b\}=2%
\delta^{ab}$. Therefore the quantum states are $|\alpha,p^i>$ where $\alpha$
indicates $2^{15}$ bosons and $2^{15}$ fermions corresponding to the two
spinor representations of SO(32).

These are precisely the quantum states of the first massive level of the 11D
supermembrane, as computed in \cite{Bars:1987nr}. They also correspond to
the first massive level of the type-IIA closed string, which gives a first
signal of the relationship to the 11D M-theory as given in \cite{Bars:1995uh}%
.

The SO(10) covariant multiplets of bosons and fermions given in \cite%
{Bars:1987nr} are massive 11D states but, through the present work, they are
now being interpreted as massless in 12D. These $2^{15}+2^{15}$ states
provide a representation of the 12D constrained superPoincar\'{e} algebra or
of the unconstrained 11D nonlinear superalgebra.

We emphasize that $2^{15}+2^{15}$ are just the transverse SO(10) components
of covariant fields in 12D. By extending the tensor and spinor indices of
these states to covariant 12D indices, one should be able to identify the
SO(11,1) covariant tensors and spinors that describe the 12D massless
supermultiplet and provide a representation of the constrained superPoincar%
\'{e} algebra covariantly. In turn, by reduction from 12 to 11, these can
also be understood as 11D covariant states that correspond to the first
massive level of the supermembrane.

\section{Remarks}

We have shown that there is a sense in which the superPoincar\'e algebra in
twelve dimensions exists: it leads to a nonlinear algebra in eleven
dimensions which contains the 11D superPoincar\'e algebra and which is
interpreted as the off-shell superparticle in eleven dimensions, as given in
Eq.(\ref{11D1}-\ref{11D3}). This algebra necessarily contains a
noncommutative spacetime in eleven dimensions $\left[ X^{\mu },X^{\nu }%
\right] \neq 0.$ The algebra is represented on the quantum states of the
first massive level of the 11D supermembrane, or first massive level of the
type-IIA closed superstring which has a close relationship to 11D M-theory.

The 11D aspect is an indication of M-theory, while the 12D aspect hints a
possible relationship to F-theory \cite{vafa} or S-theory \cite{Bars:1996ab}.

In the enlarged space including the extra dimension the algebra is the
standard superPoincar\'e algebra, but with covariant constraints. We found
that some of the constraints were unfamiliar. For example, in 11 dimensions
the constraints $P^2=0$ and $\not{P}Q=0$ are standard, but the constraints $%
\Delta^{M_1M_2\cdots M_p}=0$ for $p=3,4,\cdots,11$ were not noticed before.
The story is similar in the twelve dimensional case, with the new
constraints $\Delta^{m_1m_2\cdots m_p}=0$ for $p=3,4,\cdots,12$. These new $%
\Delta$'s commute with the translation generators $P$ and
supersymmetry generators $Q$. In general they would be related to
additional quantum numbers that label the representation. But in
our case we have a special representation in which the additional
quantum numbers all vanish. In this representation all constraints
are solved explicitly by writing the algebra as a nonlinear
algebra in one lower dimension. The representation space that
realizes the algebra is the massless particle in the higher
dimension, which is also interpreted as the off-shell particle in
one lower dimension.

It is clear from the explicit discussion in ten and eleven
dimensions that the same kind of analysis can be applied in any
number of dimensions. This could shed light on the meaning of
supersymmetry in dimensions higher than eleven\footnote{%
Without giving the details we state the result for the case of 9
to 10 dimensions, which could be done as an exercise by the
reader. The structure is similar. Namely, the nonlinear nine
dimensional off-shell N=1 superalgebra is related to the type IIA
constrained superPoincar\'e algebra in ten dimensions. The
representation space consists in the massless states of type IIA
supergravity. Similarly some aspects of the case of 4 to 5
dimensions is briefly discussed at the end of section 3.}.

\section*{Acknowledgements}

We are very grateful to Sergio Ferrara for pointing out references
\cite{S}\cite{PT}. I.B. is in part supported by a DOE grant
DE-FG03-84ER40168. He is grateful to the CERN TH-division for
hospitality while this work was performed. C.D. is supported in
part by the Turkish Academy of Sciences in
the framework of the Young Scientist Program (CD/T\"{U}BA--GEB\.{I}%
P/2002--1--7). A.P. and B.Z. are supported in part by the DOE under contract
DE-AC03-76SF00098 and in part by the NSF under grant 22386-13067.


\end{document}